\documentclass[preprint,12pt]{elsarticle}

\usepackage{amssymb}

\journal{Physica E}

\begin{document}

\begin{frontmatter}



\title{Anomalous magnetotransport and cyclotron resonance of high mobility magnetic 2DHGs in the quantum Hall regime }


\author[1, 2]{U. Wurstbauer}
\author[1]{S. Knott}
\author[1]{C. G. v. Westarp}
\author[1]{N. Mecking}
\author[1]{K. Rachor}
\author[1]{D. Heitmann}
\author[2,3]{W. Wescheider}
\author[1]{W. Hansen}

\address[1]{Institute for Applied Physics,University of Hamburg, Hamburg, Germany}
\address[2]{Institute for Experimental and Applied Physics, University of Regensburg, Germany}
\address[3]{Solid State Physics Laboratory, ETH Zurich, Zurich, Switzerland}

\begin{abstract}
Low temperature magnetotransport measurements and far infrared transmission spectroscopy are reported in molecular beam epitaxial grown two-dimensional hole systems confined in strained InAs quantum wells with magnetic impurities in the channel. The interactions of the free holes spin with the magnetic moment of 5/2 provided by manganese features intriguing localization phenomena and anomalies in the Hall and the quantum Hall resistance. In magnetic field dependent far infrared spectroscopy measurements well pronounced cyclotron resonance and an additional resonance are found that indicates an anticrossing with the cyclotron resonance. 

\end{abstract}


\end{frontmatter}

\begin{figure}[h]
\centering
\includegraphics[width=0.5\textwidth]{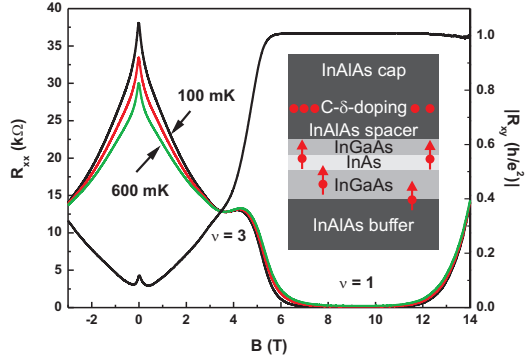}
\caption{Hall resistance trace $R_{xy}$ at $T$ = 100 mK (black) and longitudinal magnetoresistance resistance $R_{xx}$ along the [110] direction for $T$ = 100 mK (black), 400 mK (red)  and 600 mK (green) for sample B. Quantum Hall plateaus and SdH oscillations at filling factors 1 and 3 and strong localization effect in the low-field region are visible. The inset shows the layer sequence of the active QW structure of sample B.}
\label{fig:03}
\end{figure}

Manganese (Mn) doping of semiconductors such as III-V, II-VI or germanium are studied intensely in the last year due to its prospect for spin-polarized charge carrier injection and spintronic device applications. The interaction of magnetic moments with the charge carriers spin leads to a variety of phenomena like ferromagnetic ordering, localization effects and metal-insulator transition \cite{1,2,3,4}. Further, the interaction of magnetic moments with the spin of itinerant charge carriers in magnetic two-dimensional charge carrier systems amplifies the spin splitting ($\Delta$E$_{S}$) and results  in striking features in the quantum Hall regime \cite{5,6}. There, the highly degenerate Landau levels (LL) are broadened by the presence of (magnetic) disorder and has a "mobility edge" separating localized states in the tail region from extended states in the central region of each LL. The crossing of such LL of opposite spin directions has been reported in magnetic two-dimensional electron gases (M2DEGs) in II-VI heterostructures \cite{3,6}, in non-magnetic 2DEGs in InAs/InGaAs quantum wells \cite{7} and non-magnetic two-dimensional hole systems in GaAs/AlGaAs heterostructures \cite{8}. 
Here, we focus on magnetic two-dimensional hole gases (M2DHGs) in the presence of magnetic ions, which feature localization phenomena magnetic fields lower than 3 T \cite{9,10} and intriguing properties at finite magnetic fields in the quantum Hall regime. 2DHGs additional co-doped with few Mn ions supplying magnetic moment of 5/2 that can interact with the spins of the free holes have been investigated with low-temperature magnetotransport measurements and far-infrared (FIR) spectroscopy to gain information about the valence band structure.
\begin{figure}[h]
\centering
\includegraphics[width=0.5\textwidth]{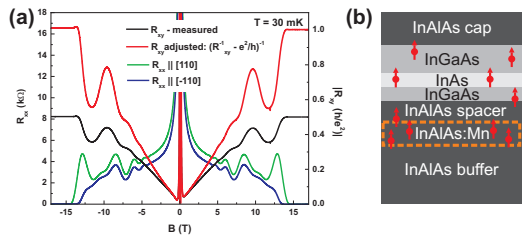}
\caption{(a) Longitudinal resistance $R_{xx}$ along the anisotropic [110] (green) and [$\overline{1}$10] (blue) direction and Hall resistance trace $R_{xy}$ as measured (black) at $T$ = 30 mK reveal Shubnikov-de Haas oscillations with vanishing resistance and quantum Hall plateaus. In the Hall resistance $R_{xy}$ additional minima and a bending to lower resistance values are visible. Adjusting $R_{xy}$ by $(R_{xy}^{-1} - e^{2}/h)^{-1}$ (red) gives a straight line with the slope corresponding to the two-dimensional hole density $p$. (b) Layer sequence of the active QW structure of sample A. }
\label{fig:01}
\end{figure}
The 2DHGs are confined in a 20 nm thick InGaAs quantum well (QW) containing an asymmetric embedded additional compressively strained InAs channel with InAlAs barriers. The doping layers are 5 nm separated from the QW. The samples are grown by molecular beam epitaxy (MBE) on semi-insulating (001) GaAs substrates. Due to the high In content of 75 $\%$ in the active region a metamorphic buffer for strain relaxation and adjustments of the lattice constant is grown between the substrate and the QW structure \cite{10}. Two different types of QW structures have been investigated in this study. Sample A is inverted modulation doped with a 7nm thick InAlMnAs layer. Peculiarities at the MBE growth lead to an asymmetric broadening of the doping layer and to a significant amount of Mn ions in the QW. The Mn concentration in the doping layer is less than $1 \cdot 10^{20}$ cm$^{-3}$ and in the QW below $1 \cdot 10^{18}$ cm$^{-3}$. A scheme of the layer sequence is depicted in figure \ref{fig:01}(b). The free holes in sample B are provided by the non-magnetic acceptor carbon (C) which is deposited after the QW in a delta doping layer. The whole active layers are co-doped with Mn ions. The concentration is estimated from flux calibration of the effusion cell to be less than $7 \cdot 10^{17}$ cm$^{-3}$. A scheme of structure B is inserted in figure \ref{fig:03}. 
\begin{figure}[h]
\centering
\includegraphics[width=0.5\textwidth]{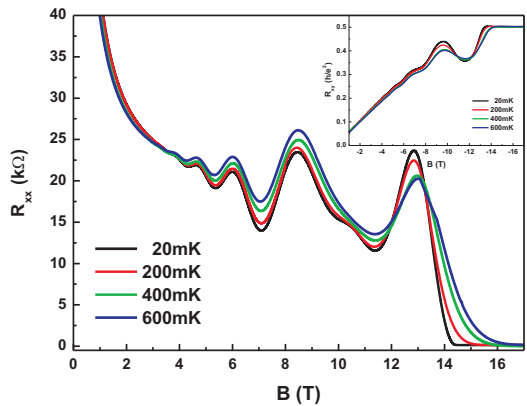}
\caption{Longitudinal magnetoresistance traces $R_{xx}$ of sample A along the [110] direction for temperatures between $T$ = 20 mK and $T$ = 600 mK. The weakly visible maximum at about $B$ = 10.5 T disappears above $T$ = 400 mK and simultaneously the maximum around $B$ = 13 T shifts to higher magnetic field values. In the inset the corresponding Hall resistance traces $R_{xy}$ are plotted. The maximum at $B$ = 9 T and the minimum at $B$ = 12 T decrease with increasing temperature. The quantum Hall-plateau around $B$ = 15 T occurs exactly at 0.5 $h/e^{2}$ for all temperatures.}
\label{fig:02}
\end{figure}
Magnetotransport measurements were carried out on L-shaped Hall bars aligned along the [110] and [$\overline{1}$10] directions that are 1000 $\mu$m long and 200 $\mu$m wide. Alloyed InZn is used to prepare ohmic contacts.  The low-temperature measurements were carried out in a $^{3}$He/$^{4}$He dilution refrigerator in the dependence of the temperature with low-frequency lock-in technique and with a bias current of 100 nA. FIR transmission spectroscopy was performed at fixed magnetic fields up to $B$ = 12 T at $T$ = 1.6 K.  Transitions between LLs were probed at frequencies between 12 cm$^{-1}$ and 300 cm$^{-1}$ corresponding to energies between 1.5 meV and 37.5 meV. A detailed description of experimental details and analysis of FIR absorptions spectra is for example given in the recent Ref. \cite{11}.\\
Figure \ref{fig:01}a) exhibits magnetotransport measurements of the Hall resistance $R_{xy}$  (black) and the longitudinal resistance $R_{xx}$ along the [110] (green) and the [$\overline{1}$10] (blue) directions on sample A at $T$ = 20 mK. The distinct difference in the magnetoresistance between the orthogonal $\langle 110 \rangle$ directions can be explained by cross-hatched morphology of metamorphic grown heterostructures due to the In concentration modulation \cite{12}. Positive Hall coefficient, distinct quantum Hall plateaus in $R_{xy}$ and Shubnikov-de Haas (SdH) oscillations in the longitudinal resistance $R_{xx}$ demonstrate transport in a high mobile 2DHG. The huge increase of the resistance in the low-field region is caused by strong localization of the itinerant holes due to the interaction with the localized magnetic ions inside the QW \cite{9,10}. A noticeable feature around $B$ = 11 T and a significant bending appear in the Hall resistance trace $R_{xy}$. Assuming the existence of a parallel conducting channel to be responsible for the bending of the Hall slope we found that, surprisingly, the deviation from a straight slope can be adjusted by subtraction of one universal conductance value from the Hall conductance. This implies an additional conducting layer with the same hole concentration as the 2DHG channel. This assumption is contradictory to the fact that the longitudinal resistance vanishes for higher field values \cite{13}.  In figure \ref{fig:01}(a) the curve of $(R^{-1}_{xy} - e^{2}/h)^{-1}$ is plotted (red trace) demonstrating a straight line with one gradient corresponding to the two-dimensional hole density $p$ at lower fields. The feature between filling factor 2 and 1 still remains and is accompanied by an additional oscillation in the longitudinal resistance. The anomalous signatures in $R_{xy}$ and $R_{xx}$ decrease with increasing temperature as depicted in figure \ref{fig:02}, where $R_{xy}$ (inset) and $R_{xx}$ along [110] is plotted for temperatures between $T$ = 20 mK and $T$ = 600 mK. The vanishing longitudinal resistance for both transport direction in the whole temperature range excludes a parallel conducting channel, e.g. the InAlMnAs layer, to be the reason for the bending of $R_{xy}$ \cite{13}. The hole density calculated from the adjusted Hall slope is confirmed by the $(1/B)$ periodicity of the SdH oscillations and is $p$ = $5.0 \cdot 10^{11}$ cm$^{-2}$. From the temperature dependent damping of the amplitude of the SdH oscillations the effective mass of the holes was determined to be $m^{*}_{dos, [110]}$ = 0.36 m$_{0}$ along the [110] direction and $m^{*}_{dos, [\overline{1}10]}$ = 0.25 m$_{0}$ along the [$\overline{1}$10]. The effective mass from the cyclotron resonance was determined to $m^{*}_{CR}$ = 0.16 m$_{0}$ at $B$ $>$ 6 T, lower than that determined from magnetotransport \cite{14}. The resonance signal in the FIR measurements done on sample A at $T$ = 1.6 K was very weak pointing towards high amount of localized hole states in each LL. Similar behavior was found for a couple of Mn modulation doped InAs QWs with an inverted doping layer.\\
For comparison, we investigate sample structure B, where the itinerant holes are provided by nonmagnetic C and the whole structure is weakly co-doped with Mn-ions. From low-temperature magnetotransport measurements positive Hall coefficient, quantum Hall plateaus without additional features and bending are observable in $R_{xy}$ and in $R_{xx}$ SdH oscillations. The hole density consistently was determined to $p$ = $2.9 \cdot 10^{11}$ cm$^{-2}$. The negative magnetoresistance in the low-field region is less distinct compared to sample A due to a reduced amount of Mn in the channel leading to less localization. At higher magnetic field values a well pronounced plateau in $R_{xy}$ and minima in $R_{xx}$ at filling factor 1 is visible, weak signatures at filling factor 3 and no signatures at filling factor 2 and 4.
\begin{figure}[h]
\centering
\includegraphics[width=0.5\textwidth]{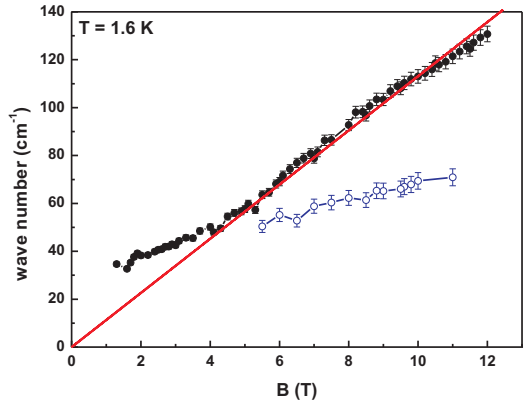}
\caption{Dispersion of the cyclotron resonance on sample B at $T$ = 1.6 K. At higher fields two resonance features are resolved resulting in two dispersion branches.}
\label{fig:04}
\end{figure}
Here, the effective mass from the temperature damping was not determinable. Only a weak temperature dependency was found.  Again, the vanishing longitudinal resistance for all investigated transport directions and temperatures exclude a parallel conducting channel \cite{13}. FIR measurements show well defined cyclotron resonance with an effective mass $m^{*}_{CR}$ = 0.08 m$_{0}$.  At low magnetic fields a deviation of the dispersion of the cyclotron resonance and at higher fields an additional resonance was found, as demonstrated in figure \ref{fig:04}. This behavior can be caused by carrier localization as seen in magnetotransport experiments or can be interpreted as a crossing or anti-crossing of two resonance signals that can not be resolved due to damping of the resonance signal at lower field.\\
In contrast to the reported findings on 2DHGs with magnetic impurities in the channel, no anomalous high-field transport behavior is observed in Mn modulation doped QW without Mn in the channel \cite{16}. This indicates that the anomalies are caused by the interaction of the spins of the itinerant holes with the localized moments of 5/2 $\mu$B provided by the Mn ions. The non-parabolic valence band structure with a complex LL fan resulting in crossing of LL with opposite spin directions and the interaction with magnetic moments of the Mn ions are similar to the findings reported in M2DEGs \cite{5,6}. Consequences of this behavior and/or the effect of repulsive and attractive scattering centres as reported in 2DEGs in GaAs heterostructures \cite{16} may lead to an explanation for the outstanding features found in 2DHGs with a significant amount of Mn in the channel. Comparison of sample A and B indicates that higher amount of Mn-ions in the channels seems to increase the intriguing high-field transport phenomena.\\
In conclusion, we have investigated magnetic 2DHG with a significant amount of Mn in the channel by low-temperature magnetotransport measurements and FIR spectroscopy. Anomalies in the high-field quantized transport properties and in the dispersion of the cyclotron resonance were found. The magnetic ions in the channel seem to be responsible for these effects.\\ \\
We gratefully acknowledge M. Habl for helpful discussions. This work was supported by the DFG via Grant No. SFB 689 and 508 and the excellence initiative "LEXI".


\begin{thebibliography}{00}
\bibitem{14}
We found similar differences in the effective mass determined from magnetotransport $m^{*}_{dos}$ and from cyclotron resonance $m^{*}_{CR}$ in two-dimensional hole systems in high-mobility GaAs/AlGaAs heterostructures.
\end{thebibliography}
\end{document}